\documentstyle[12pt]{article}
\def\beqa{\begin{eqnarray}}
\def\eeqa{\end{eqnarray}}
\def\beq{\begin{equation}}
\def\eeq{\end{equation}}
\begin{document}
\renewcommand{\theequation}{\thesection.\arabic{equation}}
\def\bib#1{[{\ref{#1}}]}
\begin{titlepage}
       \title{Inertial Effects on Berry's Phase of Neutrino Oscillations}
\author{S. Capozziello and G. Lambiase
 \thanks{E-mail address:capozziello, lambiase@physics.unisa.it} \\
 {\em Dipartimento di Scienze Fisiche "E.R. Caianiello",  }\\
 {\em Universit\'a di Salerno, 84081 Baronissi (Sa), Italy.} \\
 { \em Istituto Nazionale di Fisica Nucleare, Sez. di Napoli, Italy.} \\ }
\date{\today}
\maketitle
\begin{abstract}
The Berry phase of mixed states, as neutrino oscillations, is calculated in a
accelerating and rotating reference frame. It turns out to be depending on the vacuum
mixing angle, the mass--squared difference and on the coupling between the momentum of
the neutrino and the spinorial connection. Berry's phase for solar neutrinos and its
geometrical aspects are also discussed.
\end{abstract}

\bigskip

 PACS: 03.65.Bz, 1460.Pq, 95.30.Sf.

 Keywords: Foundations, theory of measurements, miscellaneous theories (including
Aharonov-Bohm effect, Berry's phase), Neutrino oscillations, General Relativity and
Gravitation.

\thispagestyle{empty}

\vfill

\end{titlepage}

\section{\bf Introduction}
\setcounter{equation}{0}

Particle interferometry has been recognized as a tool to carry out delicate measurement
of physical quantities that can be related to the difference in phase of the
interferring beams. Besides, the observation of quantum mechanical phases provides
accurate informations about quantities determining the phase shift (as for example, the
flux of a magnetic field) and allows to test the quantum behaviour of systems. A
remarkable contribution in this topic has been given by Berry in his pioneering paper.
In his work \cite{Berry}, he showed that a quantum state acquires an additional phase
(Berry's phase) $\beta^g$, which is related to the geometry of parameter space, besides
the {\it normal} dynamical phase $\phi$. He considered a non--degenerate quantum system
in an initial eigenstate of a Hamiltonian which varies adiabatically through a circuit
$C$ in the parametric space. The state evolves under the Schr\"odinger equation
remaining in each instant an eigenstate of the Hamiltonian.

Aharonov and Anandan \cite{AHAR} reformulated and generalized Berry's result by
disregarding the parameter space and considering an arbitrary, not necessarily
adiabatic, cyclic evolution in the projective Hilbert space ${\cal P}({\cal H})$, which
is the space of one--dimensional subspace (called also rays) of an appropriate Hilbert
space ${\cal H}$ (see also \cite{Simon}).

Nevertheless, in all formulations, Berry's results are based on non--relativistic
quantum mechanics and thus are not covariant. The covariant generalization of Berry
phase was derived in Ref. \cite{PAP} where, by using the proper time method \cite{Fock},
the authors showed that
 \begin{eqnarray}\label{1}
 \beta^g&=&i\oint_C
 <\tilde{\Psi}(\lambda)\vert\frac{\partial}{\partial\lambda_a}
 \vert\tilde{\Psi}(\lambda)>d\lambda_a\\
 &=&\phi+i\oint_C
 <\Psi(\lambda)\vert\frac{\partial}{\partial\lambda_a}
 \vert\Psi(\lambda)>d\lambda_a\,{,}\nonumber
 \end{eqnarray}
where $\lambda_a$ are the evolution parameters depending on the proper time, the state
$\tilde{\Psi}(\lambda)$ satisfies the cyclic condition,
$\vert\tilde{\Psi}(\Lambda+\lambda)>=\vert\tilde{\Psi}(\lambda)>$, and
$\vert\Psi(\Lambda+\lambda)>=e^{i\phi}\vert\Psi(\lambda)>$.

On the experimental side, a series of striking experiments have also been carried out
and their results provide direct or indirect evidences for the existence of such a
geometric phase \cite{SIX}.

The aim of this paper is to derive the geometrical phase attributed to the cyclic
evolution of a mixed state, as for example, neutrino oscillations. This result is
carried out in an non--inertial reference frame, i.e. for an accelerating and rotating
observer. As well known, the effects of the acceleration and the angular velocity are
relevant in interferometry experiments. In fact, by using an accelerated neutron
interferometer, Bonse and Wrobleski derived the predicted phase shift of quantum systems
\cite{bonse}. Because of the validity of the equivalence principle, one expects that
this effect occurs also in a gravitational field, as has been verified in Ref.
\cite{colella}. Besides, Mashhoon has derived a coupling of neutron spin to the rotation
of a non--inertial reference frame \cite{mashhoon} from an extension of the hypothesis
of locality. In Ref. \cite{atwood}, the neutron Sagnac effect has been found using an
angular velocity of about 30 times that of Earth. The spin--rotation and the
spin--acceleration contributions to the helicity precession of fermions has been
calculated in \cite{cai}, and recently, the inertial effects on neutrino oscillations
have been analyzed in \cite{CL}.

We find that the geometrical phase, when evaluated in a non--inertial reference frame,
does depend on the parameters characterizing the physics of neutrino oscillations, i.e.
the vacuum mixing angle and the mass--squared difference, and the ones characterizing
the background geometry, the acceleration and angular velocity in our case. Besides, a
dependence on the energy of neutrino also appear. The quantum phase shift induced by the
non trivial topology of the background space--time has, as we will see, interesting
geometrical consequences in connection to the {\it curvature} of the parameter space of
states.

The layout of this paper is the following. In Sect. 2, we review the main features of
neutrino evolution in an accelerating and rotating frame (see Ref. \cite{CL}). The Berry
phase in such a frame is calculated in Sect. 3. Discussion and Conclusions are drawn in
Sect. 4.

\section{\bf Neutrino Evolution in a Non-Inertial Frame}
\setcounter{equation}{0}

In Ref. \cite{CAR}, the neutrino phase is generalized in the following way
 \begin{equation}\label{2}
 \vert\psi(\lambda)> = \sum_{j} U_{f j}
 e^{i\int_{\lambda_0}^{\lambda}P\cdot
 p_{null}d\lambda^{\prime}}\vert\nu_j>\,{,}
 \end{equation}
 where flavor and mass indices are indicated by $f$ and by
Latin letters, respectively. $U_{f j}$ are the matrix elements transforming flavor and
mass bases, $P$ is the four momentum operator generating space--time translation of the
eigenstates and $p^{\mu}_{null}=dx^{\mu}/d\lambda$ is the tangent vector to the neutrino
worldline $x^{\mu}$, parameterized by $\lambda$; $x^{\mu}=(x^0, \vec{x})$ are the local
coordinates for the observer at the origin. We use the natural units.

The momentum operator $P_{\mu}$, used to calculate the phase of neutrino oscillations,
is derived from the mass shell condition
 \begin{equation}\label{3}
(P_{\mu}+A_{G\mu}\gamma^5)(P^{\mu}+A_G^{\mu}\gamma^5)=-M^2_f\,{,}
 \end{equation}
 where $A_G^{\mu}$ are related to the spinorial connections appearing into the
covariant Dirac equation in curved space--time \cite{WEI}, $M^2_f$ is the vacuum mass
matrix in flavor base
 \begin{equation}\label{4}
M^2_f=U\left(\begin{array}{cc}
                m_1^2 & 0 \\
                 0 & m_2^2 \end{array}\right)U^{\dagger}\,{,}\qquad
U =\left(\begin{array}{cc}
                \cos\theta & \sin\theta \\
                 -\sin\theta & \cos\theta \end{array}\right)\,{.}
 \end{equation}
$\theta $ is the vacuum mixing angle. Ignoring terms of the order ${\cal O}(A_G^2)$ and
${\cal O}(A_GM_f)$, one gets that, for relativistic neutrinos moving along generic
trajectories parameterized by $\lambda$, the column vector of flavor amplitude
 \begin{equation}\label{5}
\chi (\lambda)=\left(\begin{array}{c}
                           <\nu_e\vert \psi(\lambda)> \\
                           <\nu_\mu\vert \psi(\lambda)> \end{array}\right)
 \end{equation}
satisfies the equation
 \begin{equation}\label{6}
i\frac{d\chi}{d\lambda}=\left(\frac{M_f^2}{2}+p\cdot A_G\gamma^5\right)\chi\,{.}
 \end{equation}
 In deriving (\ref{6}), one uses the relation $P^0=p^0$ and
$P^i\approx p^i$ \cite{CAR}.

In a frame with acceleration $\vec{a}$ and angular velocity $\vec{\omega}$, the
components of $A_G^{\mu}$ are \cite{CL}
 \begin{equation}\label{7}
A_G^0=0\,{,}\quad
\vec{A}_G=\frac{\sqrt{-g}}{2}\frac{1}{1+\frac{\vec{a}\cdot\vec{x}}{c^2}}
\left\{2\frac{\vec{\omega}}{c}-\frac{1}{c^2}[\vec{a}\land
(\vec{x}\land\vec{\omega})]\right\} \,{,}
 \end{equation}
so that Eq. (\ref{6}) becomes
 \begin{equation}\label{8}
i\frac{d}{d\lambda}\left(\begin{array}{c}
                           a_e \\
                           a_{\mu}\end{array}\right)={\cal T}\left(\begin{array}{c}
                                                            a_e \\
                                                          a_{\mu}\end{array}\right).
 \end{equation}
$\sqrt{-g}$ is the determinant of the metric tensor of the accelerating and rotating
frame \cite{HEH}, $a_f\equiv <\nu_f\vert\psi(\lambda)>, f=e,\mu$ and the matrix ${\cal
T}$ is defined as
\begin{equation}\label{9}
{\cal T}=\left[\begin{array}{cc}
                -(\Delta/2)\cos2\theta  &
                (\Delta/2)\sin 2\theta-\vec{p}\cdot\vec{A}_G \\
(\Delta/2)\sin 2\theta-\vec{p}\cdot\vec{A}_G  & (\Delta/2)\cos2\theta
\end{array}\right]\,.
\end{equation}
up to the $(m_1^2+m_2^2)/2$ term, proportional to the identity matrix. Here
$\Delta\equiv (m_2^2-m_1^2)/2$. We restrict to flavors $e, \mu$, but this analysis works
also for different neutrino flavors. In order to determine the mass eigenstates
$\vert\nu_1>$ and $\vert\nu_2>$, one has to diagonalize the matrix ${\cal T}$. Using the
standard procedure, one writes the mass eigenstates as a superposition of flavor
eigenstates
 \begin{eqnarray}\label{10}
\vert\nu_1>=\cos\tilde{\theta}\vert \nu_e>-\sin\tilde{\theta} \vert \nu_{\mu}>\,{,}\\
 \vert\nu_2>=\sin\tilde{\theta}\vert \nu_e>+
 \cos\tilde{\theta}\vert \nu_{\mu}>\,{,} \nonumber
 \end{eqnarray}
where the mixing angle $\tilde{\theta}$ is defined in terms of the vacuum mixing angle
$\theta$
 \begin{equation}\label{11}
\tan 2\tilde{\theta}=\frac{\Delta\sin 2 \theta-2\vec{p}\cdot\vec{A}_G}{\Delta\cos
2\theta}\,{.}
 \end{equation}
 We note that $\tilde{\theta}\to \theta$ as $\vec{A}_G\to 0$
(i.e. $\vec{a}\to 0, \vec{\omega}\to 0$). The eigenvalues of the matrix (\ref{9}) are
 \begin{equation}\label{12}
\tau_{1,2}=\pm\sqrt{\frac{\Delta^2}{4}\cos^2 2\theta +\left[\frac{\Delta}{2}\sin
2\theta-(\vec{p}\cdot\vec{A}_G)\right]^2 }\,{.}
\end{equation}
 In the base of the mass eigenstates, we have
$\vert\psi(\lambda)>=a_1(\lambda)\vert\nu_1>+a_2(\lambda)\vert\nu_2>$, so that Eq.
(\ref{8}) assumes the form
 \begin{equation}\label{13}
i\frac{d}{d\lambda}\left(\begin{array}{c}
                           a_1 \\
                          a_2\end{array}\right)=\left(\begin{array}{cc} \tau_1 & 0 \\
                                                   0 & \tau_2 \end{array}\right)                                                       \left(\begin{array}{c}
                                                            a_1 \\
                                                          a_2\end{array}\right)\,{,}
 \end{equation}
 where $a_i=<\nu_i\vert\psi(\lambda)>, i=1,2$, and
 \begin{equation}\label{14}
\left(\begin{array}{c}
            a_1 \\
           a_2   \end{array}\right)=\tilde{U}\left(\begin{array}{c}
                           a_e \\
                          a_{\mu}\end{array}\right)
\,{,}\quad  \tilde{U}=\left(\begin{array}{cc}
                 \cos\tilde{\theta} & \sin\tilde{\theta} \\
                  -\sin\tilde{\theta} &  \cos\tilde{\theta} \end{array}\right) \,{.}
 \end{equation}
We have used the {\it adiabatic} condition $d\tilde{\theta}/d\lambda \approx 0$ in order
that (\ref{13}) is a diagonal matrix, i.e., we are neglecting the variations of
acceleration and angular velocity, with respect to the affine parameter $\lambda$, in
comparing to their magnitude. Eq. (\ref{13}) implies
 \beq\label{15}
 a_k(\lambda)=a_k(\lambda_0)e^{\alpha_k (\lambda )},\qquad
 \alpha_k(\lambda)\equiv -i\int_{\lambda_0}^{\lambda}\tau_k  d\lambda^{\prime}\,.
 \eeq
with $k=1,2$. For the initial condition $\vert\psi (\lambda_0)>= \vert\nu_e>$, the state
$\vert\psi (\lambda)>$ is
 \begin{eqnarray}\label{16}
 \vert\psi (\lambda)>&=&[\cos\theta_0\cos\tilde{\theta}e^{-i\tau_1\lambda}+
 \sin\theta_0\sin\tilde{\theta}e^{-i\tau_2\lambda}]\vert\nu_e>\\
     &+ & [-\cos\theta_0\sin\tilde{\theta}e^{-i\tau_1
\lambda}+\sin\theta_0\cos\tilde{\theta}e^{-i\tau_2\lambda}]
\vert\nu_{\mu}>\nonumber\,{,}
 \end{eqnarray}
where $\theta_0=\tilde{\theta}(\lambda_0)$ and the adiabaticity condition has been used
in (\ref{15}). Therefore, accelerating and rotating observers will experience a flavor
oscillation of neutrinos. For example, the probability to observe an electron neutrino
is readily calculated from (\ref{16}) and it is given by
 \beq\label{17}
 \vert <\nu_e\vert\psi(\lambda)>\vert^2=
 \cos^2(\theta_0+\tilde{\theta})\sin^2\alpha +
 \cos^2(\theta_0-\tilde{\theta})\cos^2\alpha\,{.}
 \eeq

\section{\bf Inertial Effects on Berry's Phase }
\setcounter{equation}{0}

In this Section, the Berry phase is calculated in an accelerating and rotating frame.
For convenience, let us invert Eq. (\ref{10})
\begin{eqnarray}\label{inverse}
 \vert\nu_e> &= &\cos\tilde{\theta}\vert
\nu_1>+\sin\tilde{\theta}\vert \nu_2>\,{,} \\
 \vert\nu_{\mu}> & = &-\sin\tilde{\theta}\vert \nu_1>+
 \cos\tilde{\theta}\vert \nu_2>\,{.}\nonumber
\end{eqnarray}
Repeating the previous analysis for the state $\vert \nu_e(\lambda)>$, we see that Eq.
(\ref{8}) can be read, formally, as a Sch\"odinger equation, the matrix ${\cal T}$
playing the role of the effective Hamiltonian. Then, the neutrino state $\vert \nu_e>$
{\it evolves} according to the relation
\begin{eqnarray}
  \vert \nu_e (\lambda)> & = & e^{-i{\cal T}\lambda}\,\vert \nu_e>
  \nonumber \\
  & = & e^{-i\tau_1\lambda}\cos\tilde{\theta}\vert\nu_1>+
  e^{-i\tau_2\lambda}\sin\tilde{\theta}\vert\nu_2>\,,
  \label{evolution}
\end{eqnarray}
being $\tau_1$ and $\tau_2=-\tau_1$, defined in (\ref{12}), the eigenvalues of ${\cal
T}$ corresponding to the eigenkets $\vert \nu_1>$ and $\vert \nu_2>$,  respectively (see
Eq. (\ref{13})). In a cyclic evolution, i.e. from $\lambda$ to $\Lambda+\lambda$, the
final state $\vert\nu_e(\Lambda+\lambda)>$ and the initial state $\vert\nu_e(\lambda)>$
(see Eq. (\ref{evolution})) do coincide, apart a phase factor, provided that
$\Lambda=-2\pi/(\tau_1-\tau_2)$
 \begin{equation}\label{18}
 \vert\nu_e(\Lambda+\lambda)>=e^{i\phi} \vert\nu_e(\lambda)>\,{,}
 \quad \phi=\frac{2\pi\tau_1}{\tau_1-\tau_2}=\pi\,{,}
 \end{equation}
Inserting Eq. (\ref{18}) into Eq. (\ref{1}) (for $a=1$, i.e. the evolution parameter
does coincide with the affine parameter), the Berry phase of an electron neutrino moving
along a cyclic path can be written as
 \beq\label{19}
 \beta^g_e=\phi+i\int_0^{\Lambda}<\nu_e(\lambda)\vert\frac{\partial}
 {\partial \lambda}\vert\nu_e(\lambda)> \,{,}
 \eeq
where $\phi$ is the dynamical phase evaluated in Eq. (\ref{18}). An explicit calculation
of (\ref{19}) yields the result
 \begin{eqnarray}\label{21}
 \beta^g_e&=&2\pi\sin^2\tilde{\theta}\\
 &=&
 \pi\left[1-\frac{\Delta\cos 2\theta}
 {\sqrt{\Delta^2\cos^22\theta+
 (\Delta\sin 2\theta-2\vec{p}\cdot\vec{A}_G)^2}}\right]\,{,}
 \nonumber
 \end{eqnarray}
where Eqs. (\ref{11}) and (\ref{inverse}) have been used. In similar way we get the
Berry phase of muon neutrino
 \begin{eqnarray}\label{22}
 \beta^g_{\mu}&=&2\pi\cos^2\tilde{\theta}\\
 &=&\pi\left[1+\frac{\Delta\cos 2\theta}
 {\sqrt{\Delta^2\cos^22\theta+
 (\Delta\sin 2\theta-2\vec{p}\cdot\vec{A}_G)^2}}\right]\,{,}
 \nonumber
 \end{eqnarray}
As arises from Eqs. (\ref{21}) and (\ref{22}), $\beta^g_{e, \mu}$ depend on the vacuum
mixing angle, on the mass--squared difference and energy of neutrinos, and on the
parameters characterizing, through the spinorial connection, the background geometry:
the acceleration and angular velocity. Besides, the quantum shift phase induces also a
dependence on the energy of neutrinos. Note that $\beta^g_e+\beta^g_{\mu}=2\pi$.

A direct or indirect evidence of the Berry phases (\ref{21}) and (\ref{22}) is very
difficult to achieve and how a quantum interferometry experiment could reveal it goes
beyond the aim this paper. We just analyze some interesting consequences of the above
results.

\begin{description}

\item[-] If the condition
 \beq\label{24}
 \Delta\sin 2\theta-2\vec{p}\cdot\vec{A}_G=0
 \eeq
holds, then Eqs. (\ref{11}), (\ref{21}) and (\ref{22}) imply $\tilde{\theta}=n\pi/2$,
$n=0,1,2,\ldots$, so that
 \beq\label{25}
 \beta^g_e=0 \quad \mbox{and} \quad \beta^g_{\mu}=2\pi\,.
 \eeq
Furthermore, in the rotating frame with angular velocity $\omega$ and acceleration zero,
Eq. (\ref{24}) reduces to the form
 \beq\label{26}
 \vert m_2^2-m_1^2\vert\approx \frac{4 E_{\nu}\omega}{\sin 2\theta}
 \eeq
where $E_{\nu}$ is the energy of ultrarelativistic neutrinos, $E_{\nu}\approx \vert
\vec{p} \vert $. Let us analyze Eq. (\ref{26}) for solar neutrinos and in the case in
which the observer is comoving with the Earth, i.e. its angular velocity is $\omega \sim
7\cdot 10^{-5}$rad/sec. Results of the mass--squared difference, calculated by using
(\ref{26}) for  typical values of the solar neutrino energies and vacuum mixing angle,
are reported in the Table. The agreement with the experimental data comes from neutrinos
with energy varying in the range $10\div 60$MeV. In this range in fact, we find a
mass--squared difference of the order $10^{-12}\div 10^{-10}$eV$^2$ for vacuum mixing
angle $10^{-1}\leq \sin 2\theta \leq 1$ (see also \cite{CL}).

At the present, there is a strong evidence in favor of oscillations of neutrinos and of
their non--zero masses. Such results have been found in different experiments: solar
neutrino experiments \cite{HKE}, atmospheric neutrino experiments \cite{SKK}, and the
accelerator LSND experiment \cite{LSND}. Recent reports indicate that the best fit in
favor of solar neutrino oscillations are obtained for the following case \cite{ALL1}
 $$
 \vert m_2^2-m_1^2\vert \simeq (0.5\div 1.1)\cdot
 10^{-10}\mbox{eV$^2$}\,{,}\quad \sin^2 2\theta\simeq 0.67\div
 1\,{, }
 $$
in very good agreement with our results summarized in the Table.

\item[-] If the momentum of the neutrino $\vec{p}$ is
perpendicular to the spinorial field $\vec{A}_G$, then the coupling term vanishes and
Eq. (\ref{21}) and (\ref{22}) reduce to
\begin{equation}\label{perp}
  \beta^g_e=2\pi \sin^2\theta\,,\quad  \beta^g_{\mu}=2\pi
  \cos^2\theta\,,
\end{equation}
i.e. the mixing angle $\tilde{\theta}$ coincides with the vacuum mixing angle $\theta$.
Of course, when the angular velocity and the acceleration of the frame are zero, we
recover trivially Eq. (\ref{perp}) and results of Ref. \cite{BLA}, obtained for
Minkowskian background. Eqs. (\ref{21}) and (\ref{22}) represent a generalization of the
results obtained in Ref. \cite{BLA}.

\item[-] For ultra-relativistic neutrinos and high values of the
acceleration and angular velocity, so that the coupling term is much more greater than
the mass--squared difference, the Berry phase (\ref{21}) and (\ref{22}) assume the
following values: $\beta^g_e\approx \pi$ and $\beta^g_{\mu}\approx \pi$.

\end{description}

Extension of these results can be done, invoking the equivalence principle, to
stationary gravitational fields, last ones responsible of the quantum phase shift. The
previous analysis applies also to quantum system involving boson mixing, as the
oscillations of $K^0--\bar{K}^0$.

\section{\bf Discussion and Conclusions}

In this paper, the covariant generalization of the Berry phase proposed in Ref.
\cite{PAP} has been applied to problems involving mixed states, as neutrino
oscillations. We have calculated the Berry phase of the electric and muon neutrinos in a
cyclic evolution with respect to an accelerating and rotating reference frame and we
have found that these phases do depend on the vacuum mixing angle, the mass--squared
difference and on the coupling between the momentum of neutrinos and the spinorial
connection.

The condition of no--mixing expressed by Eq. (\ref{24}), which diagonalizes the {\it
mixing} matrix (\ref{9}), implies that the Berry phase of the electron and muon
neutrinos assume the values equal to $0$ or $2\pi$, respectively. Such a condition has
been analyzed for observers comoving with the Earth, so that $\omega\sim 10^{-5}$rad/sec
and $\vert\vec{a}\vert=0$. In this particular case, the numerical values of the
mass--squared difference, calculated by using Eq. (\ref{26}) (see Table), are compared
with ones of the recent experimental data, showing a very good agreement. It is worth to
note that the inertial effects on the Berry phase, hence on the neutrino oscillations,
induce a dependence on the orientation of the rotating observer (the direction of the
angular velocity) with respect to the direction of the neutrino momenta. It implies
hence a dependence on the zenith angle recently discussed in the framework of the
atmospheric neutrino flux \cite{ALL1}.

A direct or indirect measurement of inertial (or gravitational) effects on the Berry
phases related to the neutrino particles is at the moment very difficult, considering
also the fact that the values of the mass--squared difference of neutrinos and the
vacuum mixing angle are till now open issues. Nevertheless, the Berry phases (\ref{21})
and (\ref{22}) have interesting implications from the geometrical point of view. As
already said in the Introduction, the Berry phase is attributed to the holonomy in the
parameter space. It appears when a quantum state is parallelly transported in the
parameters space around a close path. The final quantum state is rotated, with respect
to the initial one, by an angle which can be related to the connection (the Berry
connection) ${\cal A}=<\tilde{\Psi}\vert d \vert \tilde{\Psi}>={\cal A}_ad\lambda^a$
appearing in Eq. (\ref{1}), being $d=(\partial/\partial\lambda_a)d\lambda^a$ the
exterior derivative in the parameter space. Besides, one can also define a curvature
(the Berry curvature) as the field strength of ${\cal A}$, ${\cal F}=d{\cal
A}=(d<\tilde{\Psi}\vert)\wedge (d\vert\tilde{\Psi}>)$.

The existence of this phase, associated with the cyclic evolution, is universal in the
sense that it is the same for the infinite number of possible motions along the curves
in the Hilbert space, which project to a given closed curve in the projective Hilbert
space of the rays, and for the possible (effective) Hamiltonians, which rule the
evolution of the state along these curves.

In a Minkowski space--time, the Berry phase is related only to the vacuum  mixing angle:
The quantum vector state is rotated of the angle $\theta$ when parallelly transported
along a close curve, so that, as discussed above, it can be associated to the curvature
in the space of states (see Eq. (\ref{perp}) and Ref. \cite{BLA}).

In a non--inertial reference frame, new geometrical features arise, as one can see from
Eqs. (\ref{21}) and (\ref{22}):

\begin{description}

\item[a)] The parallel transport along a close curve
contains also the contribution due to the spinorial connection $A_G^{\mu}$, which
induces a quantum shift phase, hence an additional {\it rotation} in the parameter space
of the quantum vector state. The total rotation is given by the (mixing) angle
$\tilde{\theta}$, so that the Berry curvature turns out to be related both the vacuum
mixing angle and to the spinorial connection, i.e. to the non trivial geometry of the
background space--time.

\item[b)] Due to the coupling term, $\vec{p}\cdot\vec{A}_G$, the geometrical phases
(\ref{21}) and (\ref{22}) show a dependence on the energy of the neutrinos, loosing in
such a way the character of universality: different values of the neutrino energies
correspond to different Berry's phase.

\end{description}

Such a new geometrical setting follows from the generalization of the quantum mechanical
phase (\ref{2}) in which, as suggested by Stodolsky \cite{stodolsky}, the metric field
appears in the definition of the scalar product. The dependence of the geometrical phase
on the energy of particles and on the parameters characterizing the background geometry
is a common aspect of the behaviour of a quantum particle in a certain class of four
dimensional stationary space--time, as tubular matter source, slowly moving mass current
and spinning cosmic string \cite{burges,bezerra,pierri}. In the last case, for example,
it is shown that the {\it gravitational} geometrical phase of a quantum particle with
energy $E$ moving around a spinning cosmic string (with angular velocity $J$) is $\sim
JE$ \cite{pierri}.

As final comment, we stress that the previous results holds also for stationary
gravitational field, as an obvious consequence of the validity of the equivalence
principle. Then, the understanding of the issues concerning the influence of the gravity
on the Berry phase are of interest in view of theoretical arguments which try to
construct general relativity as a gauge theory and, in particular, to formulate the
symmetric second--rank tensor field in the framework of the unified gauge approach. In
such a way, the gravitational field should play the role of {\it electromagnetic} field,
in strict analogy with the Bohm--Aharonov effect, and its theoretical (and
phenomenological) consequence could help for a better understanding of the quantum
nature of particles evolving in non trivial geometries.

\vspace{0.3cm}

\begin{center}
Table: Estimation of $\vert m_2^2-m_1^2\vert$ given by (\ref{26}) as function of
$E_{\nu}$, $\sin 2\theta$, and fixed value of $\omega\sim 7\cdot 10^{-5}$rad/sec.
\end{center}
\begin{center}
\begin{tabular}{|c|c|c|} \hline\hline
$E_{\nu}$(MeV) & $\sin 2\theta$ & $\vert m_2^2-m_1^2\vert$(eV$^2$)
\\ \hline\hline
 1 & 1 & $10^{-13}$ \\
 1 & $10^{-1}$ & $10^{-12}$ \\ \hline\hline
 10 & 1 & $10^{-12}$ \\
 10 & $10^{-1}$ & $10^{-11}$ \\ \hline\hline
 $50\div 60$ & 1 & $10^{-10}$ \\ \hline\hline
 \end{tabular}
 \end{center}

\vspace{0.1cm}

\centerline{\bf Acknowledgments}

Research supported by MURST fund 40\% and 60\% art. 65 D.P.R. 382/80. \\ GL acknowledges
UE (P.O.M. 1994/1999) for financial support.

\end{document}